*A twist-grain boundary (TGB) phase in aqueous solutions of the DNA tetramer GTAC*

Gregory P. Smith[1], Chenhui Zhu[2], Mikhail Zernenkov[3], Guillaume Freychet[3], and Noel A. Clark[1]

[1]Department of Physics, University of Colorado Boulder, Boulder, Colorado, 80309

[2]Advanced Light Source, Lawrence Berkeley National Laboratory, Berkeley, California, 94720

[3]National Synchrotron Light Source II, Brookhaven National Laboratory, Upton, New York, 11973

*ABSTRACT*

We report the observation of a Twist Grain Boundary (TGB) phase of DNA, a striking motif of three dimensional (3D) equilibrium self-assembly of the DNA tetramer 5'-GTAC-3', a base sequence that is self-complimentary, pairing to form 4-base long, blunt-end Watson/Crick (WC) duplexes. Hydrophobic blunt ends and liquid crystal ordering enable these short duplexes to aggregate into long-DNA-like columns, even though the double helix has a break every 4 bases. A further step assembles these columns into monolayer sheets in which the columns are mutually parallel, and, finally, these sheets stack into lamellar arrays in which the column axis of each layer rotates in helical fashion, through a 60º angle with each passage to the next layer. This reorientation in a left-handed TGB helix enables each 2nm diameter WC column to be parallel to, and to partially enter, the major grooves of its neighbors.

The key role of DNA in biology stems from the selective interactions between the side groups (bases) on pairs of polymeric NA chains which direct molecular self-assembly, enabling matching sequences of large numbers of bases to package and transmit genetic information in duplex H-bonded columnar structures 1,2]. These columns are stabilized in aqueous solution: (*i*) by the selective Watson-Crick (WC) adenosine-thymine and guanine-cytosine base pairing motif; (*ii*) by polymerization of the bases as side groups on phospho-diester flexible polymer chains; and (*iii*) by the columnar stacking of the aromatic hydrocarbon nanosheets, driven by hydrophobicity, resulting from base pairing and polymerization. This mode of stabilization is sufficient to maintain the double helical structure of single column of long DNA duplex strands in the limit of small concentration, i.e., in absence of the interaction with adjacent columns.

On the other hand, the same WC columnar organization is also found monomeric or oligomeric DNA, in which only groups of $N \geq 1$ bases are polymerized, if they terminate with blunt ends or pairing overhangs which maintain the stacking. In aqueous solution blunt ends, for example, attract hydrophobically, the same pair-wise mutual interaction that gives duplex stacking in the first place. WC columns can be stable down to $N = 1$ in such systems if they form a liquid crystal columnar phase (COL) where they are packed side-by-side into a three dimensional (3D) hexagonal columnar lattice. In the COL phase, which is found in DNA polymers of all lengths, from monomers to long duplex strands, the parallel WC stacks are additionally stabilized via confinement by their neighboring columns. This confinement effect becomes more important for smaller $N$, becoming necessary to offset the weakening effect of having smaller oligomers, and therefore a larger number of effective breaks in the polymer strands.

Such an enhanced dependence on intercolumn repulsion should make details of intercolumn interaction more important at small $N$. This, combined with the increased number of internal degrees of freedom and higher entropy in a stack of short oligomers relative to these in a long-DNA stack, creates opportunities for a broader range of collective DNA behavior. We have pursued in recent years a broad study of the equilibrium self-assembly of nucleic acid oligomers in the limit where the number $N$ of bases is small, ($1 \leq N \lesssim 10$), and the DNA concentration, [c], is large, in the range 200 mg/mL $\lesssim$ [c] $\lesssim$ 1200 mg/mL [3,4,5,6,7,8,9,10,11,12,13]. At these concentrations, which are up to ~2/3 of that of dry DNA, ($[c]_{dry}$ ~1700 mg/mL [14,12]), nucleic acids self-assemble into three dimensional collective structures and phases that depend on $N$, [c], temperature, $T$, and the base sequences of the NA oligomers. Generally, as $N$ is decreased the NA osmotic pressures and therefore the concentrations required for collective self-assembly increase. In such solutions, where the oligomers are complementary or self-complementary, WC-like local organization and double helix formation can be obtained for any $N$, *i.e.* even $N = 1$, as in solutions of monomer nucleotides [12], leading to the observation of isotropic (ISO), nematic [15] (NEM) and columnar [16] (COL) phases. However, we have found that the small-*N*, nanoNA regime also proves to be a much richer area for the general exploration of novel soft-matter collective behaviors in NAs at higher concentrations. Along these lines, we have recently reported a nucleic acid phase obtained in single component aqueous solutions of the DNA tetramer 5'-GCCG-3', in which DNA spontaneously assembles into a low-density 3D mesoscale framework of WC rods, extending the horizon of nucleic acid nanofabrication to include components comprising a single molecular species, and ultra-short NAs [17]. Here we report a striking twist grain boundary (TGB) phase of the DNA tetramer 5'-GTAC-3' (GTAC).

The TGB phase was originally proposed as a possible organization of layered liquid crystal phases of chiral rod-shaped molecules [18], and first observed in a thermotropic smectic mesogen [19]. In nematic phases of chiral molecules, the molecular director, giving the local mean molecular long axis orientation, spontaneously twists into a space-filling helical precession. However, the fluid layer structure of the smectic expels both director twist and bend, allowing only splay. The TGB was envisioned as a way for smectics to accommodate chirality-preferred twist of the director, in which blocks of untwisted smectic layers were separated by walls comprising arrays of parallel screw dislocations in the smectic layering, twist grain boundaries in the smectic layering, thereby generating a twist between the director fields in adjacent blocks [18]. X-ray diffraction (XRD) provides definitive evidence for TGB ordering [20]. In the smectic TGB phases the grain boundaries are separated by blocks of untwisted layers that can range from tens of nanometers to microns thick [21]. Following this initial work on smectics, TGB structures and phases were proposed as possibilities in other LC systems, including columnar phases [22,23,24]. There have been possible twist grain boundary sightings in long DNA made on the basis of observation of arrays of crossing columns in a state of DNA crystals growing as triangular plates [25,26], or of orthogonally crossed planes of columns in a DNA columnar phase [27], but no definitive evidence, as of 1996 [28] or more recently, for a TGB phase in DNA. A bio-generated TGB columnar array of chitin fibrils is found in the stomatopod dactyl club [29].

The molecule investigated here is the DNA oligomer 5'-GTAC-3' in which the phosphodiester four-base chain is -OH terminated on both its 5' and 3' ends. Synthesis, purification details are presented in Materials and Methods. GTAC is a self-complementary palindrome forming 4-base blunt-end WC duplexes. These are ~13Å long along the column axis, comprising ~0.4 turn of a long-DNA double helix. As described above, these duplexes aggregate by hydrophobic stacking of the blunt ends to form extended locally WC-duplex columns, as sketched in *Figures 1C,D*, effectively having breaks in the two polymer strands every 4 bases. For drawing purposes *Figure 1C* shows the extended double helix in the columns as continuous at the end-to-end duplex junctions, but this is not necessarily the case. Small increments of azimuthal duplex orientation about the column axis at these junctions will change the helix pitch and, in a structure of interacting columns, environmental conditions that act to alter the azimuthal orientation of individual GTAC duplexes may alter this internal column geometry.

*Results*

Aqueous solutions of GTAC were prepared in capillaries and between glass plates and studied by polarized transmission optical microscopy (PTOM) and lab-based as well as and synchrotron-based wide and small angle x-ray diffraction (WAXS & SAXS). Of critical importance in this study was the use of synchrotron-based microbeam WAXS on the Soft Material Interfaces (SMI) x-ray microbeam line at NSLS-II. As concentration increases the kinetics of equilibration of nanoDNA phases becomes slower, making it more difficult to anneal the quality single domains needed for phase characterization. However, with samples in 50$\mu$m x 500$\mu$m flat thin wall capillaries and a 2$\mu$m x 25$\mu$m 16.1 keV x-ray beam profile, many domains of coexisting phases and alignment textures form and can be scanned in a single capillary, as shown in *Figure 2*, enabling the x-ray, and accompanying optical observation to identify essentially perfect monodomains of the various phases (*Figures 1,2*). The sample orientation typically obtained in the NA liquid crystal phases has the column axes on average parallel to the flat capillary surfaces and normal to the beam, In this geometry the periodicity of the base stacking in the columns is a clear feature in the XRD pattern of all of the LC phases, indicating the column axis (*z*) orientation in the COL phase. To illustrate achievable monodomain quality we show in *Figure 1A,B* the remarkable similarity of the WAXS pattern of a monodomain of the COL phase of columnar aggregates of the Drew-Dickerson DNA dodecamer (DD) $N = 12$ (5'-CGCGAATTCGCG-3'), to that of Rosalind Franklin's "Photo 51" XRD pattern of a shear aligned COL phase monodomain of calf thymus long DNA [2].

GTAC in solution exhibits the following series of liquid crystal phases with increasing DNA concentration: isotropic (ISO) – [350 mg/ml] – nematic (NEM) – [470 mg/ml] – deformable columnar (COL) and uniaxial columnar (COLX) – [650 mg/ml] – twist grain boundary (TGB). Transitions among these phases are first order and strongly hysteretic, such that, depending on concentration and temperature variation, coexistence of the NEM, COL, COLX, and TGB, or any two or three of these can be achieved. For example in Figure 1 the COL, COLX, and ISO coexist. Since DNA is a chiral molecule these phases must all exhibit some chiral manifestations.

Apart from the ISO, the generally obtained alignment of these phases obtained in the flat capillaries has the column axes parallel to the flat glass walls and normal to the beam. In this geometry, the WAXS shows the column internal structure, but structure along the beam direction is not observable. As a result, we supplemented the monodomain x-ray experiments with synchrotron-based powder XRD, including rapid scanning mode [30], which enabled dynamic phase transition behavior to be captured in series of scans.

We introduce the phase structures found in GTAC as follows:

•BASE STACKING – We define for the planar geometry a coordinate system with *z* along the column axis and cell wall, *y* also parallel to the cell wall, and *x* normal to the cell wall and parallel to the x-ray beam. In this geometry *q* can be parallel to the column axis direction and the scattering shows the periodicity of the base stacking. The COL phase in long DNA, the DD 12-mer, and GTAC exhibit very similar base-stacking peaks centered at scattering vector $q_x, q_y = 0$ and $q_z \cong 1.83$Å$^{-1}$, indicating a stacking periodicity of 3.4Å. This peak is diffuse, with a half-width at half-maximum (HWHM) in $q_z$ of $\delta q_r$ ~0.02Å$^{-1}$, indicating correlated periodic stacking of the bases over lengths of $\delta \xi_z$ ~50Å along the columns, and a HWHM radius in the $q_x, q_y$ plane of $\delta q_r$ ~0.25 Å$^{-1}$, approximately the diffraction limited peak width expected for a stacking of flat objects ~10Å in diameter. The base-stacking peaks in *Figure 1C,D* show that $\delta \xi_z$ is somewhat larger in the COLX and TGB. The base stacking peak decreases in intensity with decreasing concentration but is observable in all of the LC phases discussed below, indicating the presence of column segments in all of them.

•COL – The COL phase exhibits a fluid PTOM texture of a uniaxial array of columns, exhibiting the polarized light extinction brushes between crossed polarizer/analyzer (P/A) of conformal fans, observable in *Figure 2*, which indicates bend-only in-plane deformation of the column axes, typical of fluid columnar phases [16], as also seen in Refs. [5-15]. The bend curvature of the columns, dominant in the PTOM textures of the COL phase, requires free sliding of columns along the column axis direction (*Figures 1A,B*). If, by chance, the column axis were to be normal to the cell wall and parallel to the x-ray beam (only occasionally observe [3]) then the COL phase WAXS pattern would be an array of resolution limited spots on a hexagonal lattice, as expected from the hexagonal column packing of uniaxial columnar phase. [3] These spots are localized to the $q_x, q_y$ plane indicative of the translational symmetry along *z* of the time-averaged phase structure of the COL phase along the columns.

The typical COL alignment and the only one in the rest of this discussion is planar with the column axes parallel to the cell wall, as is all of the cases in *Figure 1*. Scattering from the column packing in planar alignment is potentially visible to the left and/or right of the beam in the (horizontal) $q_z = 0$ plane of *Figure 1A-C*, showing up when the orientation of the hexagonal lattice about *z* is such that the planes in which the columns that are closest-packed are normal to the cell wall, i.e. that one pair of the smallest-*q* reciprocal lattice vectors are in the $q_y, q_z$ plane. The diffuse ring at a radius near that of the hexagonal lattice indicated the presence of fluctuations or heterogeneities to short range local hexagonal columnar ordering, in all of the above phases.

•NEM – The nematic phase is a disordered COL phase of short columnar segments which retain a degree of nematic orientational order of their column axes. Structural fluctuations smear the COL hexagonal spot pattern into a diffuse ring, circularly symmetric about *z*. and having a diameter $q_N$ comparable to that of the neighboring COL phase from remnant column side-by-side correlations. The NEM phase shows the most dramatic effects of chirality: the typical chiral nematic (cholesteric) LC oily streak textures with Grandjean lines; and iridescent Bragg-like reflection of visible light when the pitch of the chiral helix is on the scale of hundreds of nanometers. The base stacking peak is observable in the NEM phase.

•ISO – The ISO phase The isotropic phase is a disordered NEM phase in which the columnar segments are shorter, fewer, and have lost their orientational ordering.

•COLX – The COLX phase was produced by evaporative concentration of the sample, with the mesophase obtained after one week of equilibration exhibiting large domains that, unlike the COL phase, could be rotated into extinction across the entire domain simultaneously, indicating uniform birefringence, a high degree of alignment, and more rigid order than present in the COL. There do not appear to be subdomains of different birefringence in the COLX POTM texture, indicating uniaxial and therefore hexagonal packing of the columns structure. This phase thus has elastic resistance to bend, which must come from a interaction of the structural periodicity along a column with that of its neighbors, preventing the free sliding of columns along the column axis direction required for bend curvature, producing a 3D-crystal-like hexagonal lattice. In the WAXS the strong reflections of the COLX indicate that the column axis is parallel to the cell walls and those of a planar aligned hexagonal lattice with the planes in which the columns are closest-packed are normal to the cell wall, i.e. that one pair of the smallest-$q$ reciprocal lattice vectors are along $q_y$, producing the bright spots at small $q$. The columnar hexagonal lattice fundamental $q = 0.249$ Å$^{-1}$ gives the spacing $b = 25.3$ Å between planes of closest-packed columns which is equivalent to an equilateral triangle unit cell side of $a = 29.2$ Å. This period has been seen to vary with the 2D lattice spacing becoming smaller with increasing DNA concentration. Along the column axis, there is the strong base stacking peak at $q_z = 1.87$ Å$^{-1}$ (3.364 Å) matching the conventional B-form stacking period of 3.4Å. While the strong characteristic diffuse scattering from the helix observable in *Figure 1B* in the oligomeric CGCGAATTCGCG COL phase, is essentially identical to that in *Figure 1A* in long DNA, it is absent in the GTAC COLX pattern in *Figure 1C*. There are, however, other weak scattering features in *Figure 1C*: (*i*) An on-axis peak at $q_z = 0.4676$ Å$^{-1}$, corresponding to a period along *z* of 13.4 Å, which is the length of the blunt-end duplex GTAC segments in the aggregated column. This peak is sharp along *z*, indicating coherent periodic ordering over ~10 duplex lengths along the columns, The peak width in $q_y$ is similar to that of that of the base-stacking as would be expected for the stacking of duplexes; (*ii*) Weak additional, diffuse lines of scattering in $q_y$ and $q_z$, extended along $q_y$, at $q_z = 0.194$ Å$^{-1}$. Three possible origins of these features suggest themselves: •They may be the remnant of the X pattern for scattering from very short helix segments; •They may indicate transient correlation in the local orientation and position in GTAC duplexes in different columns; or they may be a pretransitional feature of approach to the TGB phase, as they are coincident in $q_y$ and $q_z$ position with the Miller index 10 and 11 Bragg peaks in the TGB phase hexagonal lattice (*Figure 2D*). Otherwise, the major scattering features of *Figure 1C* show the COLX phase to be made of columns that, apart from the base stacking, are rendered essentially featureless by orientation and position fluctuations of their short duplexes.

•TGB – Following such evaporation runs, some domains in the flat capillaries showed no measurable birefringence in PTOM (*Figures 2B,3*), initially suggesting an isotropic phase at high concentration  However, microbeam XRD from these black areas produced a breathtaking six-fold symmetric scattering pattern (*Figures 1D,2D*). As indicated by the familiar DNA base-stacking peaks at $q = 1.87$ Å$^{-1}$, this phase shows three sets of base stacked DNA columns, each with the columns running parallel to the cell walls, but with different orientations in the plane of the cell wall (azimuthal angle $\varphi = 0°, 60°, 120°, 360°$), and therefore crossing each other at a 60° angle between axes, implying that they at different distances from the cell wall. In PTOM there was some variability in the degree of extinction of these dark domains, indicating weak birefringence (*Figure 2D*). Subsequent microbeam XRD and PTOM observation showed that in the weakly birefringent areas the XRD pattern had lost six-fold symmetry, sometimes exhibiting only two-fold symmetry or none at all. Analysis showed that in the birefringent six-fold areas the axis of the columns of a given $\varphi$ made some finite angle with the cell wall, meaning that the different $\varphi$ sets made different angles relative to the beam direction, the *x* axis in *Figures 1,2*, breaking symmetry and also producing birefringence. There is evidence for such a small column tilt of this nature in the TGB WAXS pattern of *Figure 1D* where it can be seen that around the ring of second-order hexagonal diffraction spots there is a systematic variation in peak intensity. Such tilts are strictly accidental, depending on how the domain grew in, but the strongest alignment tendency is for the columns to be parallel to the cell wall, with many domains found where the tilt effect is small and does not substantially affect the scattering pattern, as in *Figure 1D*.

Apart from the ISO, the general alignment of these phases obtained in the flat capillaries has the column axes parallel to the flat glass walls and normal to the beam. In this geometry, the WAXS shows the column internal structure, like base stacking and column arrangement in the $q_y,q_z$ plane. Structure along the beam (*x*) direction of planes of columns is not observable. As a result, we supplemented the monodomain x-ray experiments with synchrotron-based powder XRD, including rapid scanning mode [30] obtained in the COLX-TGB coexistence region [*c*] > 500mg/ml, where the sample was heated to isotropic and cooled to a temperature, ending back in the COLX phase, at which time a transition phase front to the TGB passed through, evolving the scattering pattern from COLX to TGB, behavior captured in a series scans, shown in *Figure 3*.

Referring to F*igures 1C,D*, the COLX and TGB the columnar arrangements are each characterized by two principal distances:

COLX – •$b \approx 25.3$Å, the center-to-center spacing between the planes of closest-packed columns ($q_b \approx 0.248$ Å$^{-1}$ from the radius locating the spots in *Figure 1C*). •$a = b(2/\sqrt{3}) \approx 29.2$Å, the center-to-center spacing between columns in the closest-packed planes of columns in the hexagonal lattice (also the edge length of the 60º equilateral parallelogram unit cell of the hexagonal lattice); Unit cell area = $ab$ = 25.3Å x 29.2Å. The COL and COLX column spacings are understood to be variable, decreasing with increasing osmotic pressure, such that $a,b \propto 1/[c]^{½}$.

TGB – •$l \approx 27.8$Å, the center-to-center spacing in *y*,*z* plane between columns in each planar column layer ($q_l = 0.226$ Å$^{-1}$ from the radius of the ring of spots in *Figure 1D*); •$h \approx 21.0$Å, the center-to-center spacing along *x* between the centers of the column layers; Effective unit cell area = $lh$ = 27.8Å x 21.0Å. In the TGB phase the *l* spacing appears quite fixed while the *h* spacing varies with sample concentration as $h \propto 1/[c]$. Other relevant TGB lengths are: •$P = 3h \approx 63$Å, the pitch along *x* of the TGB helix; •$A \approx l(\sqrt{3}/2) \approx 32.1$Å, the distance along any column between crossing points by columns in the next layer along *x* (also the length of the lattice unit cell of the TGB hexagonal lattice in the *y*,*z* plane). Using these spacings, we calculate the COLX DNA concentration in *Figure 2C* of [*c*] = 393 mg/mL, and the TGB DNA concentration in *Figure 1D* of [*c*] = 506 mg/mL.

*Discussion*

The WAXS patterns in *Figures 1A,B* make the direct comparison of the scattering from long DNA [2] and the blunt-end stacked DNA, here the 12mer 5'-CGCGAATTCGCG-3', where the latter is -OH terminated at both ends. The 12mer is in the COL phase [31], as is the long DNA sample [28]. In this case, due to the column sliding disorder of the COL phase, the scattering pattern is that from a single column, apart from the hexagonal column lattice equatorial peaks at $q_y = \pm 2\pi/b$, $q_z \approx 0$. The two images are remarkably similar which would seem to indicate that either: (*i*) the WC helix is the same in the two cases, meaning that it is continuous at every blunt-end juncture in the 12mer stacks; or (*ii*) the pattern in *Figures 1A,B* come from coherent diffraction only over distances less that the 12mer length, i.e., a single turn of the helix, with coherence lost at larger separations because of fluctuations of the helix, with the result that breaks in the helix at longer separations have little effect on the 12mer pattern. In this case the photo 51 X pattern would be the scattering from a single helix pitch, averaged over all of its azimuthal orientations around the column axes. Case (*ii*) is consistent with atomistic simulation showing that non-phosphorylated blunt end junctures tend to form significant breaks in the phase of the helix [32,33,34]. Classically, the Photo 51 diffraction pattern is understood to be a special case of generalized helical diffraction. The structure factor for a smooth helical wire winding around the z-axis with radius $r$ and helical period $P$ can be determined by solving the integral [35]:

$$T\left(R,\varphi,\frac{n}{P}\right) = \int_0^P \exp\left[2\pi i \left\{Rr \cos\left(2\pi \frac{z}{P} - \varphi\right) + \frac{nz}{P}\right\}\right] dz$$

Where the structure factor $T$ is cylindrically symmetric in q-space as defined by a radial coordinate $R$ and an angular coordinate $\varphi$ and integrated in real space over z from the origin to one helical pitch. Critically, one helical pitch contains sufficient information to produce this structure factor. And, this structure factor is non-zero only on layer lines in $q_z$ where $n$ is some integer in a ratio defined by the helical pitch $n/P$. This integral solves with Bessel J functions [35]:

$$T\left(R,\varphi,\frac{n}{P}\right) = J_n(2\pi Rr) \exp\left[in\left(\varphi + \frac{1}{2}\pi\right)\right]$$

On each layer line $n$, the intensity from the structure factor is therefore proportional to a square of the nth Bessel function $J_n$. As such, the Bessel functions on the layer lines $n$ display a first maximum that widens in radius from the column axis given increasing $q_z$ and simultaneously decreases in intensity with increasing $n$. This results in a layered 3D cone-like structure that projects upward and downward away from the origin surrounding the axis of the helix. A 2D slice at some angle through the $q_z$ axis gives an "X"-like shape where the first maximum out from the origin for $J_1$ has the greatest intensity in the series [2,35].

Photo 51 is further modulated by the inclusion of (*i*) a layered stack of DNA bases with a z-axis spacing of ~3.4Å, giving discretization of the smooth helix period that is nearly commensurate with the 10.4 base pairs stacked per helical turn and (*ii*) a second backbone helix in the duplex that is out of phase with the first by a physical displacement of about 9 to 12 Å along the z-axis [2]. Together, these modulations account for layer line omissions at $n = 4$ and the distortion of those occurring above $n = 5$ up to the base-stacking peak at ~1.85 Å$^{-1}$ [1,2,35]. Consistent with the discussion above, for the long DNA used to produce Photo 51 (Figure 2A), the diffraction pattern comes directly from $T\left(R,\varphi,\frac{n}{P}\right)$, implying that neighboring columns are decorrelated along the column axis direction while in the 2D hexagonal LC lattice, which does not affect the pattern. For the COL phase of the DNA 12mer, where the diffraction pattern strongly resembles that of Photo 51, the structure factor must be mainly preserved for an object that experiences double strand breaks just longer than the period of the B-form helix (every 12 bases). For the 12mer DNA, the peak signifying that double strand break every 12 bases is at a q-space position so close to the origin that it's behind the beam stop and not located where it could interfere with the pattern's similarity to Photo 51, which is all at larger $q$. Given that Photo 51 itself deviates from the expected $J_0$ layer line [2], potential distortions to the ideal helix, the structure factor of which only depends on one helical turn anyway, apparently aren't spaced closely enough together to break the similarity between the Photo 51 and 12mer DNA diffraction patterns.

These considerations can also be applied to *Figure 1C*. With GTAC, the column break occurs in the backbone at every fourth base, which appears in the diffraction pattern at $(0, 0, 0.4676 \text{ Å}^{-1})$. The coherent integration giving $T$ is now over only a range of length $\sim P/3$, which is then averaged over the helix phase and duplex orientational and positional variations. As a result the coherent scattering from the GTAC $N = 4$mer internal structure is relatively much weaker (varying as $N$), and otherwise coherent scattering from neighbors reduced by frequent breaks in the helix. In the COLX bend is suppressed so there must be positional correlations preventing the relative sliding of columns along the z axis, which may give the linear diffuse feature along $q_z = 0.194$ Å$^{-1}$, peaked at finite $q_y$. Otherwise, the only scattering are the $q_z = 0$ spots at finite $q_y$, coming from the hexagonal column lattice, and the base stacking peak. Thus, the columns appear featureless in the x-ray pattern, apart from the base stacking, which brings us to *Figure 1D*.

It is useful to consider the TGB pattern in *Figure 1D* as a superposition of three COLX patterns as in *Figure 2C*, rotated through 0°, 60°, and 120°, respectively, about the $x$ axis, to give the six-fold symmetric ring of spots (hexagonal Miller index 10) from the side-by side packing of columns in the layers parallel to the $y,z$ plane. In the TGB these columnar side-by-side packing peaks are sharp, but at a somewhat smaller $q$. Even making this adjustment, however, it is clear that such a superposition will not show the most striking feature of the pattern in *Figure 1D*, namely its array of six next-order (hexagonal Miller index 11) diffraction spots. This is because the dominant scattering features in *Figure 1C* show the COLX phase to be made of columns that, apart from the base stacking, are rendered essentially featureless by orientation and position fluctuations of the short duplexes. Visual inspection of *Figure 1C* shows that the conceived simple superposition of the three scattering of arrays of featureless columns will have no (11) scattering peaks. Thus, the (11) scattering must be coming from the structuring of the columns occurring in the assembling of the TGB phase when the layers of columns are stacked and choose to cross. This layer stacking produces a peak in the in the TGB powder scattering (*Figure 3*), from which we obtain the layer spacing along $x$ of $h = 21.0$Å, a column layer spacing that is remarkably small compared to the smallest spacing of column layers in the COLX ($b = 25.2$ Å), or to the spacing within the $y,z$ layer planes in the TGB ($l = 27.8$ Å). This small spacing along $x$ of the crossed columns requires their interpenetration at the crossings, which is facilitated by manipulating the columns' secondary structure through appropriate orientational and positional sequence of the GTAC duplexes enabling crossing columns to mutually fit. The columns are no longer featureless. This accommodation, periodic along the columns with period $A \approx l(\sqrt{3}/2) \approx 32.1$Å, produces the (11) scattering spots in the TGB reciprocal lattice, such that the proposed superposition accounts for the major features of the scattering in *Figure 1D*. This, along with the single peak giving the $h = 21$Å periodicity along $x$, indicates that there are no layers of different structure (*e.g.*, COL or COLX layers) between the grain boundaries, *i.e.*, every layer of columns along x separates two grain boundaries. It is worth also pointing out that these 11 peaks, while clean and strong, have no detectable higher harmonics, meaning that the modulation of the electron density along a column is sinusoidal, suggesting a smooth variation of structure along the column. Picking the column axes in one of the layers then a set of 10 and 11spots are along the $q_z = 0.194$ Å$^{-1}$ line, looking like a Bragg version the *z* axis in the TGB a given column direction the

Placing columns in a lattice of crossings introduces a collective local external periodic potential on the rotators in each column, introducing the possibility of long range ordering, if mutual arrangements of duplex orientations at the crossings of sufficiently low energy can be found. Within the WC duplexes the helical twist along a column (*z*) axis is right-handed, but this makes the twist through the column, along say the *y* axis, left-handed, saying more generally that the opposite sides of the columns are chirally related. This makes it virtually certain that the TGB stacking of the column layers along *x* will be helical. But a macroscopically helical TGB stacking will establish the same chiral geometry at every column crossing, making it virtually certain that the column are also helical. Thus, it may not be just coincidence that the spacing of crossings along columns in the TGB is comparable to the WC helical pitch, but rather that having $A \sim \mathcal{P}$ is an effective way to organize the columns in response to a the same chiral kick at every crossing. We model this situation by assuming that the column structure is a slightly compressed long DNA continuous helix [the long DNA (LDNA) model] in order to explore the fitting of the columns at the crossings, as sketched in *Figure 1D*. Here we show a stack along *x* of four columnar layers in which the GTAC orientations in a single column are adjusted to a pitch $\mathcal{P} = 32.1$Å to match the lattice parameter *A*, with a 60° rotation between each pair. Each 60° crossing assembly of pairs of columnar layers is a twist grain boundary (TGB) in the *y,z* plane, and the stacking of such boundaries along *x* into a periodic array creates a twist grain boundary "TGB phase" [18].

The LDNA TGB model creates a helical reorientation of the columnar layers along *x*, which helix can be either left handed or right handed. Because of the inherent chirality of the duplexes, such left handed and right handed TGBs will be in no sense mirror symmetric structures. The x-ray pattern in *Figure 1D* enables us to say with certainty that, while the intra-column helix elements in the GTAC duplexes have the typical right-handed WC helix, the experimental TGB helix along *x* in GTAC in the LDNA model is the left handed one, as also shown in *Figure 1D*. In a packing of parallel columns the backbone ridges on the columns approach those of their neighbors in an "X" crossing geometry. In the left-handed TGB helix the backbone ridges on a column end up running nearly parallel to the axes of the neighboring columns, enabling these columns to fit some distance into its major groove  This geometry is clearly evident in the sketches of *Figure 1D*, which additionally show that in the left handed TGB stacking of layers 1, 2, 3, 4 , say at 0º, 60º, 120º, and 360º,layers 1 and 4 have their columns not only parallel in the *y,z* plane, but their column axes also project parallel to *x* to exactly the same place on the *y,z* plane (*Figure 1D*).  This produces the strong first order reflection ring of spots in the six-fold pattern corresponding to column plane separation, *l*.  In the right hand helix the columns of layers 1 and 4 are also parallel, but layer 1 columns project halfway between layer 4 columns on the *y,z* plane, a structure in which the first order diffraction spots in the six fold pattern will cancel, and their second harmonic will be bright. Absence of the second harmonic eliminates the possibility of a right hand TGB helix in the LDNA model.

*Materials and Methods*

<u>*Sample Preparation*</u> – Short oligomers of GTAC and 12mer 5'-CGCGAATTCGCG-3' were synthesized by the Caruthers method as in (refs) using an Akta Oligopilot 100 and purified using isopropanol precipitation. Oligomer samples were then dried by lyophilization and stored for later use.

Samples used for powder scattering SAXS/WAXS were typically loaded into 1.5 mm diameter borosilicate glass melting point capillaries with one end presealed. While keeping track of the masses added by scale, solid granular DNA was gently tamped into the end of capillary and centrifuged to collect the powder at the bottom of the capillary. A set volume of water was then injected into the capillary by a 1 $\mu$L micropipettor and centrifuged to unite solid with liquid. By tracking loaded masses of these components, we could estimate the sample concentration which enabled us to systematically target particular liquid crystal phases. These capillaries were then pull-sealed in the flame of oxygen-propane torch to permanently trap the water and thermocycled to 90° to mix the water with the DNA. We examined visible features of the resulting mesophases using PTOM in a jig containing an index matching oil to help correct for the rounded surfaces of the capillary wall.

Samples used for microbeam XRD probing were prepared in flat capillary tubes 50$\mu$m x 500$\mu$m in dimension. Unlike the cylindrical capillaries described previously, these capillaries are not robust to centrifugation and must be loaded in a different manner. Here, the solid DNA was dissolved in water at an ISO concentration to remove order and promote flowing. An open end of the capillary was then immersed in the DNA sample and ISO DNA allowed to crawl into the tube by capillary effect. Before the ISO liquid fully filled the tube, the immersed end was then switched to a heavy fluorinated oil that happens to be volatile. The layer of fluorinated oil also draws into the tube by capillary effect and mostly lifts the ISO phase ahead of it, allowing the DNA mixture to be moved away from the mouth of the capillary. With the DNA plug sufficiently centered in the middle of the capillary, the tube was lifted from the fluorinated oil and the oil allowed to evaporate off. We then underwent short cycles of forced water evaporation by shifting the sample briefly into a vacuum produced by a roughing pump. Achieving the correct mesophase by these means included some guess-work since the material weights were difficult to judge in a controllable manner. As water evaporates out, a very strong DNA concentration gradient is formed at the water-air interface in the capillary, giving very high-order phases at the interface and lower concentration to ISO phases internally and essentially locking in the water. Upon achieving a perspective concentration, the capillary was flame-sealed at either end and allowed to sit at room temperature for a protracted period of up to a week. In these conditions the DNA concentration gradients tend to equilibrate and uniform mesophase textures emerge. Some of these textures even tend to orient during evolution by interactions with the walls of the capillary. The careful cleaning process and flame-sealing of these capillaries produce samples that are vacuum compatible, retaining water stably without evaporative loss. Because of their flat walls, these samples are considerably more amenable to PTOM then cylindrical specimens, but the throughput is much slower given the guesswork involved in forming target textures.

*X-ray diffraction* – Cylindrical specimens used for powder scattering XRD were shot in open air through the thick-wall borosilicate by synchrotron illumination at ALS beamline 7.3.3. The beam here was 10 keV energy and the beam spot size was about 300 $\mu$m. The thickness of the walls limit utility with weaker X-ray fluence. Because of their stability, these samples could be ramped through a thermocycle by hot-stage while in the X-ray beam without losing water by evaporation from the sample and thereby retaining mesophase behaviors at a desired concentration.

Flat capillary specimens used for microbeam XRD at NSLS-II beamline 12D, were examined almost exclusively at room temperature, but within a strong vacuum to support observation of detailed scattering and removing interference by atmosphere on dim scattering features. The beam here was 16.1 keV and the spot size 2$\mu$m x 25$\mu$m. Clear diffraction was almost certainly facilitated by the relatively thin 50$\mu$m thick aspect of the capillary, eliminating 3D complexity in the specimen; not optimally thin for optical examination, but significantly more selective than the 1 mm cylindrical capillaries.

X-ray Data were processed using the Nika 2D SAS package available for Igor Pro. Diffraction patterns were also examined using custom code via the Python PIL package and by ImageJ to recover certain line scans for detailed feature positioning.

*Model Building* – 3D models of the LC phase were built using the widely available 3DS Max graphics software package. The column model was predicated to be a smooth helix for ease of assembly. Column footprint was assembled on a 2 nm diameter circle, with the base-pair represented by a simple ellipsoid shape less than 0.15 nm in thickness offset from the center of the circle and straddling between two edges. Further ellipsoids were positioned at the tips of this base-pair to schematically represent the backbone of the helix with these ellipsoids tipped in a direction that mimics the precession of the helix. The helix parameters were than assembled directly from parameters measured in the XRD experiments described above: a 3.364Å translation with a 37.704° right-handed twist. Taking this translation and twist to the single base model and repeatedly transforming copies of that base as many times as desired, produces a smooth helix with a 9.6 base/turn pitch and a 2nm diameter with the rough profile of B-form DNA. A single helix model could then be placed into layered arrangements that were stacked with 60° twists to give the TGB helix described in the text and seen in Figure 2D. Columns in each layer were then rotated around their unique column axes independently of one-another until arrangements emerged that were favorable to meshing between neighboring layers.

*Acknowledgments*

This research was supported in part by the NSF Biomaterials Program under Grant DMR-1611272, by NSF MRSEC Grant DMR-1420736 to the University of Colorado Soft Materials Research Center, and by the NIH/CU Molecular Biophysics Training Program to GPS. This research used the microfocus Soft Matter Interfaces beamline 12-ID of the National Synchrotron Light Source II, a U.S. Department of Energy (DOE) Office of Science User Facility operated for the DOE Office of Science by Brookhaven National Laboratory under Contract No. DE-SC0012704. We also acknowledge the use of beam line 11.0.1.2 of the Advanced Light Source at the Lawrence Berkeley National Laboratory supported by the Director of the Office of Science, Office of Basic Energy Sciences, of the U.S. Department of Energy under Contract No. DE-AC02-05CH11231.

*Figures*

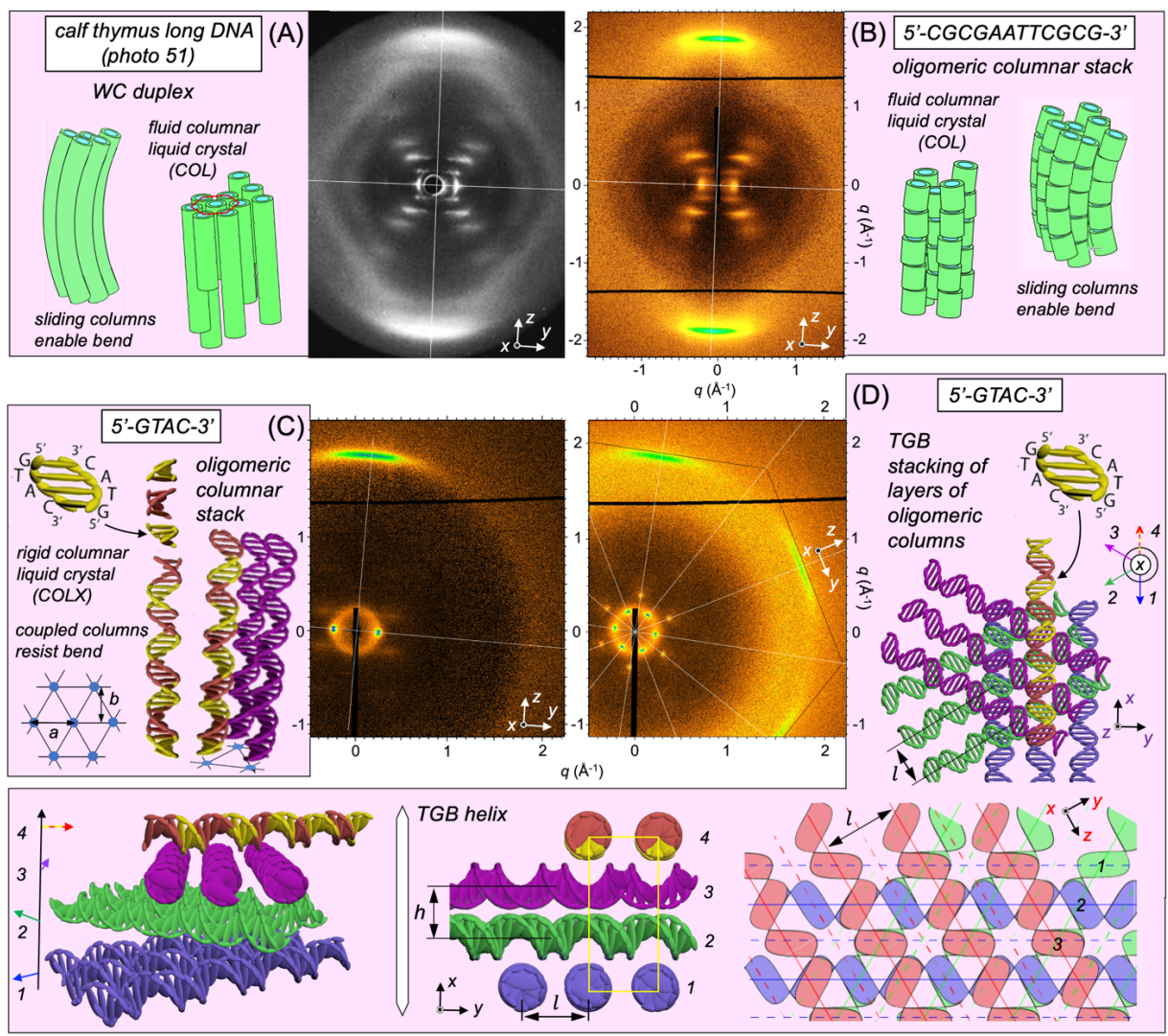

*Figure 1*: (*A*) Scattering and structure of the COL phase of long calf thymus DNA (Photo 51 [2]). (*B*) Scattering and structure of the COL phase of the -OH terminated DNA dodecamer 5'-CGCGAATTCGCG-3'. (*C*) Scattering and structure of the COLX phase of the -OH terminated DNA tetramer 5'-CTAC-3'. (*D*) Scattering and structure of the TGB phase of the -OH terminated DNA tetramer 5'-CTAC-3', with several different motifs for visualizing the TGB structure. The red *x,y,z* axes are for the red layer in the 2D layering motif.

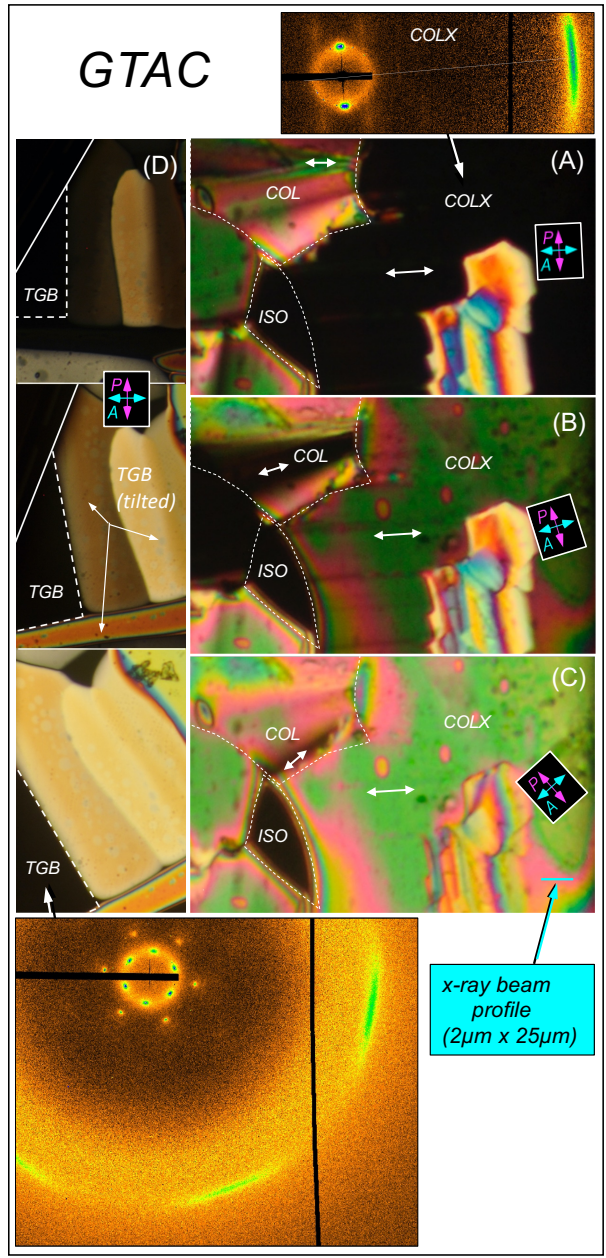

*Figure 2*: 50 μm thick single domains of isotropic (ISO), fluid columnar (COL), ordered columnar (COLX), and twist grain boundary (TGB) phases of duplex GTAC in areas of thinwall flat capillaries, imaged in PTOM and showing the x-ray diffraction of the indicated domains, the latter collected with NSLS SMI x-ray microbeam illumination (2μmx25μm in cyan). PTOM crossed P/A extinction indicates the mono-domain phases: rotating extinction brushes = COL; completely extinguishing only for certain P/A orientations = uniaxial COLX; completely extinguishing for all P/A orientations = ISO or TGB.

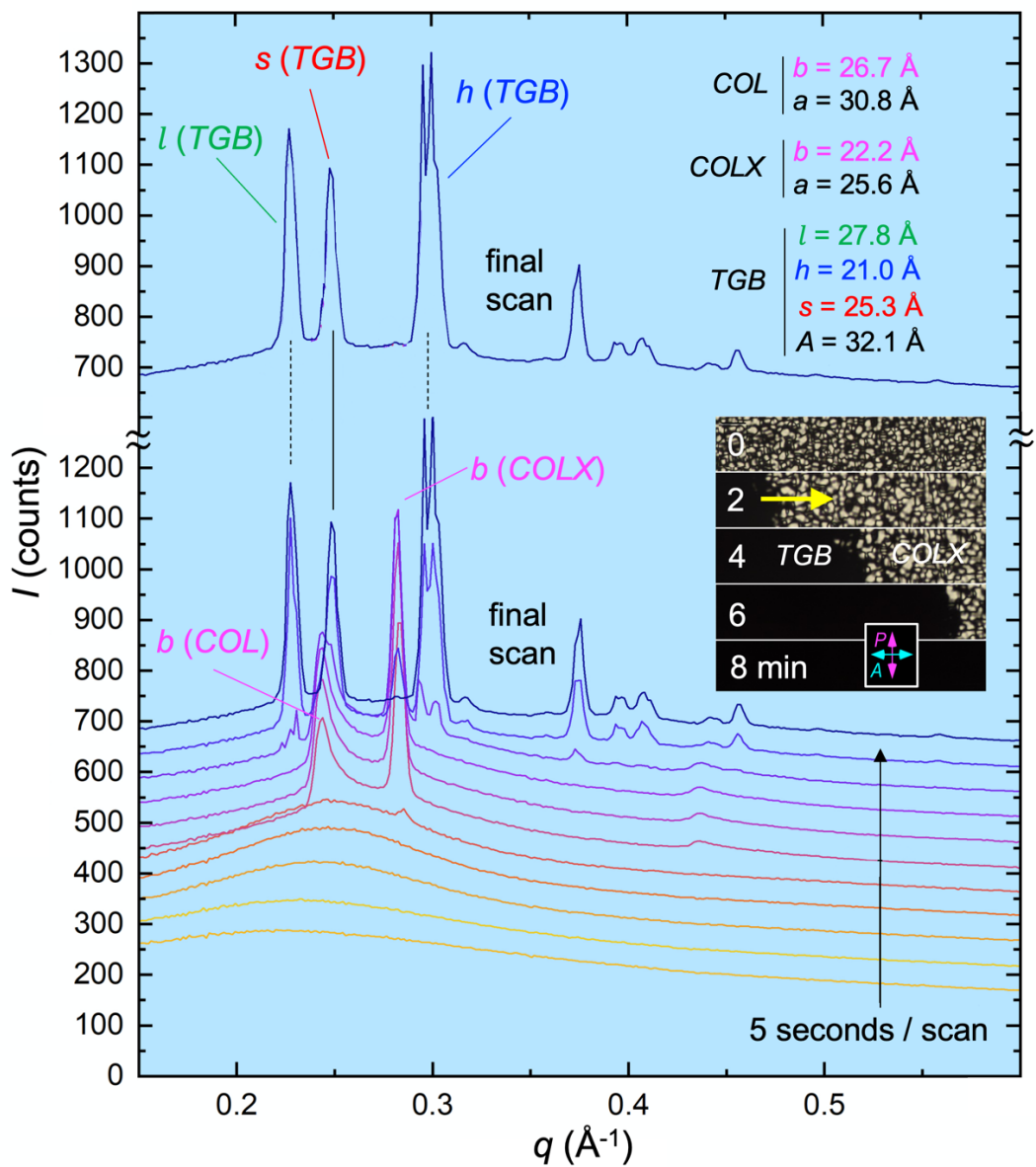

*Figure 3*. Powder-average XRD of phase-change dynamics during a thermal quench of a GTAC DNA solution filled in a capillary at $[c] = 610$ mg/ml. Inset: PTOM showing a phase transition between the COLX and TGB phases. The sample is heated to the ISO phase, cooled back into the NEM phase, and held in temperature until the ordered phases grew back in. They did so with distinct moving phase fronts. The image shows the COLX/TGB front separating the isotropic TGB, dark between crossed polarizer and analyzer, from the birefringent COLX. Rapid scanning at the Advanced Light Source, beamline 7.3.3, Berkeley enables XRD probing as the front moves through the x-ray beam producing the corresponding set of scans plotted. Colored font indicates spacings obtained from the peaks in this plot, and black ones are calculated. The NEM phase has a broad peak from side-by-side positioning of transient column segments. The sharper peaks

mark the growth of the hexagonal lattice of columns in the COL and COLX phases, giving the spacing (*b*) of the closest packed layers of columns, where onlycthose volumes in the sample of hexagonal phase having the column axis *z* normal to the Ewald sphere at *q* = 0 contribute to the scattering. The COLX coexists initially with the TGB, but then abruptly disappears. The final scan is plotted also separately for clarity, exhibiting three TGB peaks, now associated with those volumes in the sample where a TGB *x*,*y* plane, illustrated in *Figure 1D*, is tangent to the Ewald sphere. The unit cell then observed is indicated in yellow in *Figure 1D*: the spacing (*l*) of the columns in the layer planes normal to the TGB helix axis; the spacing (*h*) along *x* of these layers of columns; the structure period (3*h*) along *x*. The three principal TGB peaks in the final scan give the dimensions of the corresponding rectangular reciprocal lattice vectors of the yellow unit cell. The leftmost peak is $q_y = g_y = g_{01} = (2\pi/l) = 0.227$Å$^{-1}$. The second peak is from the unit cell diagonal $q_s = g_{11} = [g_x^2 + g_y^2]^{½} \equiv (2\pi/s) = 0.248$Å$^{-1}$, where $g_x = g_{10} = (2\pi/3h) = 0.100$Å$^{-1}$. The third peak is $q_x = 3g_{10} = (2\pi/h) = 0.299$ Å$^{-1}$, which is the first peak appearing with increasing *q* if *q* is along *x*, a result of the electron density of a layer being independent of its orientation when projected onto the *x* axis, making the period = *h* for *q* along *x*, and canceling the peaks at $q_x = g_{10}$ and $2g_{10}$.